%% file: main.tex
\documentclass[doublecolumn,10pt] {IEEEtran}
\usepackage{cite}
\usepackage{amsmath,amssymb,amsfonts}
\usepackage{graphicx,color}
\usepackage{textcomp}
\usepackage{xcolor} 
\definecolor{elsevierblue}{HTML}{2fb2e2} 

\usepackage[colorlinks=true,
            linkcolor=elsevierblue,  
            citecolor=elsevierblue,  
            urlcolor=elsevierblue    
           ]{hyperref}
\usepackage{enumitem}
\usepackage{multirow}
\usepackage{titlesec}
\usepackage{academicons}
\usepackage{xcolor}
\newcommand*\emptycirc[1][1ex]{\tikz\draw (0,0) circle (#1);} 
\newcommand*\halfcirc[1][1ex]{%
  \begin{tikzpicture}
  \draw[fill] (0,0)-- (90:#1) arc (90:270:#1) -- cycle ;
  \draw (0,0) circle (#1);
  \end{tikzpicture}}
\newcommand*\fullcirc[1][1ex]{\tikz\fill (0,0) circle (#1);} 
\usepackage{subcaption} 
\usepackage{placeins}
\usepackage{booktabs}
\usepackage{tikz}
\usetikzlibrary{arrows.meta, positioning}
\usetikzlibrary{mindmap, shadows}
\usepackage{tabularx}
\usepackage{dblfloatfix} 

\usepackage[linesnumbered,ruled,vlined]{algorithm2e}

\definecolor{orcidlogocol}{HTML}{A6CE39}

\def\BibTeX{{\rm B\kern-.05em{\sc i\kern-.025em b}\kern-.08em
    T\kern-.1667em\lower.7ex\hbox{E}\kern-.125emX}}

\makeatletter
\renewenvironment{abstract}{%
  \begin{center}%
    {\normalfont\bfseries Abstract}%
  \end{center}%
  \normalfont 
}{%
  \par\vspace{0.5em}%
}

\let\oldIEEEkeywords\IEEEkeywords
\renewenvironment{IEEEkeywords}{%
  \oldIEEEkeywords
  \normalfont 
}{%
  \endlist
}
\makeatother

\begin{document}

\title{

MOSAIC: Multi-Domain Orthogonal Session Adaptive Intent Capture for Prescient Recommendations}

\author{%

\textbf{Abderaouf Bahi\textsuperscript{1\S}}~\href{https://orcid.org/0009-0003-7116-5080}{\includegraphics[height=1em]{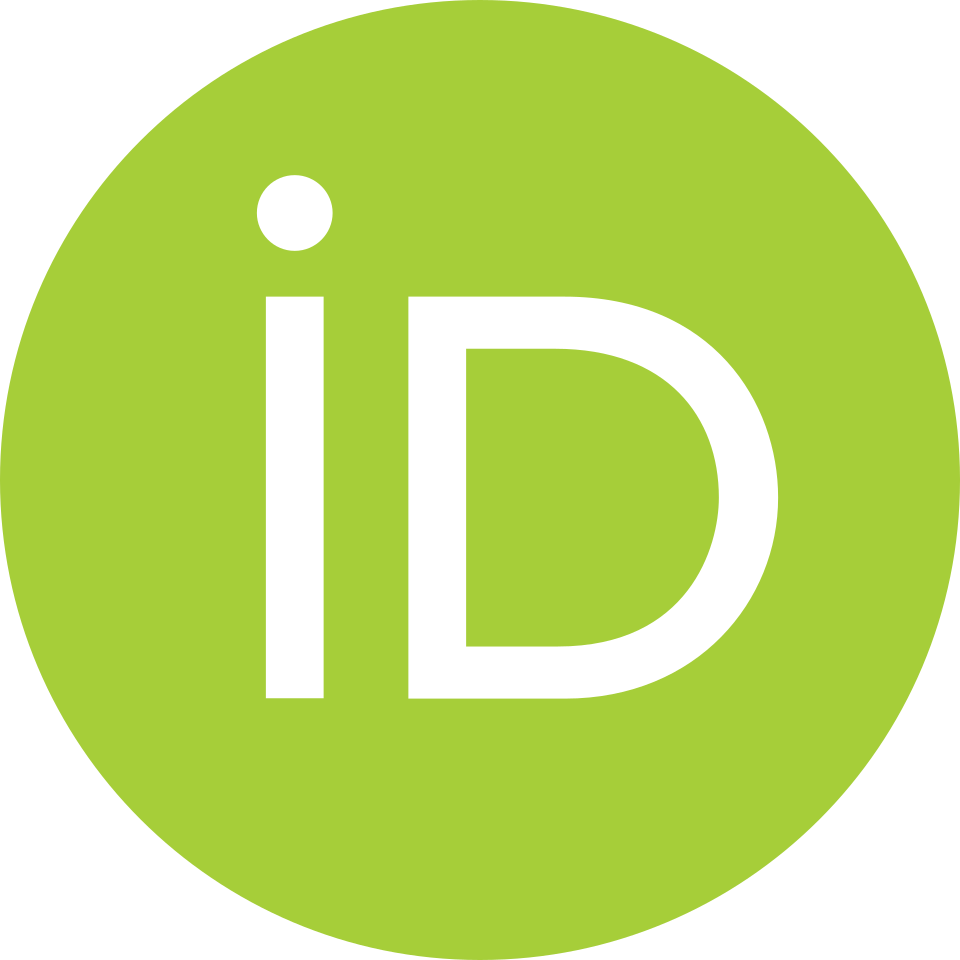}}, 
\textbf{Mourad Boughaba\textsuperscript{2}}, 
\textbf{Ibtissem Gasmi\textsuperscript{1}}~\href{https://orcid.org/0000-0002-8939-1727}{\includegraphics[height=1em]{ORCID_iD.png}}, 
\textbf{Warda Deghmane\textsuperscript{1}}~\href{https://orcid.org/0009-0007-0362-1514}{\includegraphics[height=1em]{ORCID_iD.png}}, and
\textbf{Amel Ourici\textsuperscript{3}}~\href{https://orcid.org/0000-0001-5701-8576}{\includegraphics[height=1em]{ORCID_iD.png}} \\[0.5em]
\textsuperscript{1}\small
Computer Science and Applied Mathematics Laboratory (LIMA),
Faculty of Science and Technology, Chadli Bendjedid University, 
P.O. Box 73, El Tarf 36000, Algeria \\[0.5em]
\textsuperscript{2}\small
Electro-Mechanical Systems Laboratory (LSEM), Faculty of Technology, Badji Mokhtar University, 
P.O. Box 12, Annaba 23000, Algeria \\[0.5em]
\textsuperscript{3}\small
Mathematical Modeling and Numerical Simulation Laboratory (LAM2SIN),
Faculty of Technology, Badji Mokhtar University, 
P.O. Box 12, Annaba 23000, Algeria \\[0.5em]
\textsuperscript{\S}Corresponding author: \textbf{Abderaouf Bahi} (a.bahi@univ-eltarf.dz) \\[0.5em]
}

\maketitle
\begin{abstract}
Capturing user intent across heterogeneous behavioral domains stands as a 
fundamental challenge in session-based recommender systems. Yet, existing 
multi-domain approaches frequently fail to isolate the distinct contribution 
of cross-domain interactions from those arising within individual domains, 
limiting their ability to build rich and transferable user representations. 
In this work, we propose MOSAIC, a Multi-Domain Orthogonal Session Adaptive 
Intent Capture framework that explicitly factorizes user preferences into 
three orthogonal components: domain-specific, domain-common, and 
cross-sequence-exclusive representations. Our approach employs a 
triple-encoder architecture, where each encoder is dedicated to one 
preference type, enforced through domain masking objectives and adversarial 
training via a gradient reversal layer. Representational alignment and mutual 
independence constraints are jointly optimized to ensure clean preference 
separation. Additionally, a dynamic gating mechanism modulates the relative 
contribution of each component at every timestep, yielding a unified and 
temporally adaptive session-level user representation.
We conduct extensive experiments on two large-scale real-world benchmarks 
spanning multiple domains and interaction types. The ablation study 
validates that each component—domain-specific encoding, domain-common 
modeling, cross-sequence representation, and dynamic gating—contributes 
meaningfully to the overall performance.
Experimental results demonstrate that MOSAIC consistently outperforms 
state-of-the-art baselines in recommendation accuracy, while simultaneously 
providing interpretable insights into the interplay between domain-specific 
and cross-domain preference signals. These findings highlight the potential 
of orthogonal preference decomposition as a principled strategy for 
next-generation multi-domain recommender systems.
\end{abstract}



\def\BibTeX{{\rm B\kern-.05em{\sc i\kern-.025em b}\kern-.08em
    T\kern-.1667em\lower.7ex\hbox{E}\kern-.125emX}}

\section{Introduction}
\label{sec:intro}

Session-based recommendation, which aims to anticipate a user's next 
interaction by mining preference signals from historical behavioral 
sequences, has become an indispensable component of modern online 
platforms such as e-commerce, streaming services, and content discovery 
systems \cite{sun2019bert4rec, hidasi2015gru4rec, kang2018sasrec}. 
Nevertheless, conventional single-domain models are inherently 
constrained by data sparsity \cite{sparsity2024ssl}, which amplifies 
cold-start difficulties and skews preference learning toward the dominant 
patterns of a single domain \cite{amid2025, cdsrnp2024}. To overcome 
these limitations, multi-domain sequential recommendation has emerged as 
a compelling paradigm that harnesses user behavioral signals across 
heterogeneous domains to enrich preference modeling and improve 
recommendation quality \cite{chen2024survey, zhu2025survey, homomorphic}.

Contemporary multi-domain approaches exploit both individual-domain 
sequences and cross-domain sequences that consolidate interactions 
spanning several domains \cite{cao2022c2dsr, ye2023dream, smartgrid, neom}. Individual domain sequences encode domain-specific 
behavioral patterns, whereas cross-domain sequences can amplify latent 
preference signals that remain imperceptible within a single domain 
\cite{lin2024man, xu2024llmcdsr}. For instance, a user's inclination 
toward the romance genre may appear too weak to detect in the Book domain 
alone, yet becomes prominently discernible once reinforced by correlated 
signals from the Movie domain within the cross-sequence \cite{cao2022c2dsr}. 
Simultaneously, cross-domain sequences may carry redundant cues that 
duplicate patterns already captured by individual-domain sequences, while 
other signals remain exclusively discoverable through cross-domain 
co-occurrence \cite{bian2025abxi, ma2023tricdr}. Consequently, effectively 
separating overlapping information from cross-sequences while preserving 
domain-distinctive signals is paramount for constructing more informative 
and transferable user representations \cite{cmvcdr2024, cdcl2024}.

Beyond redundancy, cross-domain sequences can surface entirely novel 
behavioral patterns absent from any individual domain in isolation 
\cite{chen2024survey}. A user who first reads a literary classic and 
subsequently watches its film adaptation exemplifies a cross-domain 
behavioral trajectory exclusively uncovered through the joint sequence 
\cite{ye2023dream, yang2024ftcf}. However, the relevance of such 
cross-domain signals is not static — it fluctuates dynamically across 
timesteps \cite{bian2025abxi, hgn2025}. A romance preference inferred 
from a cross-domain session may carry little predictive value when 
forecasting the next interaction in a fantasy-oriented or action-driven 
context \cite{tclrec2025}. This temporal variability underscores the 
necessity of an adaptive mechanism capable of modulating the relative 
influence of cross-domain and domain-specific signals at each prediction 
step \cite{hgn2025, wang2026atacdsr}.

To harness these diverse behavioral signals, prior works have leveraged 
graph neural networks \cite{wu2022gnnsurv, fedgcdr2024}, 
Transformer-based encoders \cite{sun2019bert4rec, lin2024man, sfnn}, and gating 
mechanisms \cite{hgn2025} to integrate cross-domain information into user 
representations. More recent contributions have further incorporated 
cross-sequences into their architectures \cite{bian2025abxi, ye2023dream}, 
refining domain representations through shared-attention and alignment 
strategies \cite{ma2023tricdr, zhao2025c2dsra2}. These advances 
collectively reflect the growing recognition that cross-domain sequences 
constitute a rich and transferable source of user preference signals that 
single-domain models fundamentally cannot exploit \cite{zhu2025survey, 
amid2025}.

Despite this progress, existing multi-domain sequential recommendation 
models face two persistent shortcomings: (1) insufficient disentanglement 
of overlapping information between cross-domain and individual-domain 
sequences \cite{cmvcdr2024, cdcl2024, macridvae2019}, and (2) limited 
capacity to adaptively regulate the degree of cross-domain influence at 
each timestep \cite{bian2025abxi, hgn2025}. When redundant cues are not 
properly separated, models risk capturing duplicate rather than 
complementary features \cite{filtervae2025, kgdrl2021}. Equally, 
cross-domain signals that prove informative in one behavioral context may 
become irrelevant or even counterproductive in another, degrading overall 
recommendation quality \cite{amid2025, wang2026atacdsr}.

To address these challenges, we propose MOSAIC, a Multi-Domain Orthogonal 
Session Adaptive Intent Capture framework for prescient recommendations. 
MOSAIC introduces a triple-encoder architecture that explicitly 
disentangles and dynamically integrates user preferences into three 
orthogonal components. The specific encoder captures domain-exclusive 
embeddings from individual-domain sequences, enforcing their independence 
from shared and cross-sequence information. The cross encoder learns 
cross-sequence representations that jointly model cross-sequence-exclusive 
signals and enhanced domain-common preferences through an alignment 
objective. During recommendation, a Transformer encoder processes the 
user's most recent interactions, whose outputs interact with precomputed 
specific and cross preferences via a token-level cross-attention gating 
mechanism. This gating adaptively modulates the contribution of each 
preference type at every timestep, and the resulting token-level outputs 
are aggregated into a unified session-level representation for next-item 
prediction. The main contributions of this work are summarized as follows:

\begin{itemize}
    \item We propose MOSAIC, a novel triple-encoder framework that 
    explicitly disentangles user preferences into three orthogonal 
    components: domain-specific, domain-common, and 
    cross-sequence-exclusive representations, ensuring that each encoder 
    captures complementary and non-redundant behavioral signals.

    \item We introduce a dynamic integration mechanism based on token-level 
    cross-attention gating with session-level aggregation, enabling the 
    model to adaptively regulate the contribution of each preference type 
    across timesteps and behavioral contexts.

    \item We conduct extensive experiments on two large-scale real-world 
    multi-domain sequential recommendation benchmarks, demonstrating that 
    MOSAIC consistently outperforms strong competitive baselines and 
    validating the effectiveness of both the orthogonal disentanglement 
    strategy and the adaptive gating mechanism.
\end{itemize}

The remainder of this paper is organized as follows. 
Section~\ref{sec:related} reviews related work. 
Section~\ref{sec:prelim} introduces the necessary preliminaries and 
formally defines the problem setting. 
Section~\ref{sec:method} describes the proposed MOSAIC framework in detail. 
Section~\ref{sec:exp} presents and analyzes the experimental results. 
Finally, Section~\ref{sec:concl} concludes the paper and outlines 
directions for future research.

\section{Related Work}
\label{sec:related}

Our work intersects with several active lines of research, including 
sequential recommendation, cross-domain recommendation, disentangled 
representation learning, and multi-domain user modeling. We review each 
area with emphasis on recent advances and their limitations relative to 
the proposed MOSAIC framework.

\subsection{Sequential Recommendation}

Sequential recommendation aims to model the dynamic evolution of user 
preferences from ordered interaction histories to predict future 
behaviors. Early approaches relied on recurrent neural networks to encode 
temporal dependencies in user sequences. The introduction of 
self-attention mechanisms marked a pivotal shift, with SASRec establishing 
unidirectional Transformer-based modeling as a strong baseline for 
next-item prediction. BERT4Rec subsequently extended this paradigm to 
bidirectional sequence modeling, adopting a Cloze-style masked item 
prediction objective that allows each item to attend to both past and 
future context, yielding richer user representations \cite{sun2019bert4rec}. 
More recently, intent-aware approaches have been proposed to go beyond 
surface-level behavioral patterns. TCLRec introduces a temporal-aware 
contrastive learning framework that explicitly distinguishes between 
incidental behaviors and stable user intentions, alleviating the 
limitations of random augmentation strategies in sequential modeling 
\cite{tclrec2025}. Despite their effectiveness, these methods operate 
within a single domain and suffer from data sparsity and cold-start 
challenges when user interactions are limited.

\subsection{Cross-Domain Sequential Recommendation}

Cross-domain sequential recommendation (CDSR) has emerged as a principled 
strategy to combat data sparsity by leveraging behavioral signals spanning 
multiple domains. C2DSR introduced a contrastive cross-domain framework 
that jointly models single-domain and cross-domain sequences via graph 
neural networks, establishing the importance of cross-sequence signals for 
preference learning. Building upon this foundation, several methods have 
sought to enhance cross-domain knowledge transfer through attention and 
alignment mechanisms. C\textsuperscript{2}DSRA\textsuperscript{2} proposes 
a contrastive approach with an attention-aware mechanism, explicitly 
modeling the linear relationship between target domain preferences and 
multi-domain user behaviors \cite{zhao2025c2dsra2}. ATA-CDSR addresses the 
practical limitation of non-overlapping users by constructing temporal-aware 
cross-domain sequential graphs augmented with dual attention at both node 
and domain levels \cite{wang2026atacdsr}. More recently, DASP frames the 
CDSR problem through domain-aware self-prompting, integrating large language 
models with lightweight domain adapters and meta-learned initialization to 
generate both accurate recommendations and natural language explanations 
\cite{boka2026dasp}. C3DSR proposes channel-enhanced contrastive modeling 
that extends attention to the channel dimension, capturing temporal 
contextual relationships missed by standard sequence encoders 
\cite{c3dsr2025}. Despite these advances, existing CDSR methods often 
struggle to fully separate overlapping cues between cross-domain and 
single-domain sequences, and few provide mechanisms to dynamically regulate 
cross-domain influence at each prediction step.

\subsection{Multi-Domain and Federated Recommendation}

As recommendation platforms become increasingly heterogeneous, multi-domain 
and privacy-preserving approaches have gained traction. MGCL proposes a 
multi-view graph contrastive learning framework for CDSR that simultaneously 
captures intra-domain sequential patterns and inter-domain user preference 
complementarity \cite{mgcl2025}. FedSCOPE addresses the intersection of 
federated learning and CDSR by combining offline LLM-generated semantic 
augmentation with decoupled contrastive learning under differential privacy 
constraints \cite{zhao2026fedscope}. FP2CDSR further investigates 
privacy-preserving cross-domain sequential recommendation through federated 
architectures combined with self-attention temporal modeling and feature 
mapping strategies \cite{chen2026fp2cdsr}. FairCDSR introduces a 
fairness-aware perspective, employing sequence augmentation and 
multi-interest learning to reduce disparities across demographic groups in 
cross-domain settings \cite{faircdsr2025}. These works collectively 
demonstrate the expanding scope of cross-domain modeling, yet they do not 
explicitly address the disentanglement of overlapping preference signals 
between individual-domain and cross-domain sequences.

\subsection{Disentangled Representation Learning for Recommendation}

Disentangled representation learning seeks to decompose latent user 
preferences into interpretable and independent factors, improving model 
robustness and controllability. MacridVAE pioneered macro-micro 
disentanglement in recommendation by factorizing user behavior into 
high-level intentions and low-level preferences through a variational 
autoencoder framework. CMVCDR extends disentanglement to the cross-domain 
setting by dividing users' general interest representations into 
domain-invariant and domain-specific components, combined with a 
cross-domain contrastive objective to impose additional alignment constraints 
\cite{cmvcdr2024}. Knowledge-guided approaches have further demonstrated 
that incorporating external knowledge graphs can make disentangled 
representations more interpretable and robust to data sparsity 
\cite{kgdrl2021}. More recently, Filter-VAE introduced a weakly supervised 
disentanglement strategy based on filter-based adaptive swapping, selectively 
exchanging stable latent factors to achieve cleaner and more meaningful 
separations \cite{filtervae2025}. Despite this progress, existing 
disentanglement methods for recommendation rarely address the three-way 
decomposition into domain-specific, domain-common, and cross-sequence-exclusive 
representations that is central to MOSAIC.

\subsection{Adversarial Learning and Domain Alignment}

Adversarial training with gradient reversal has been widely adopted to 
promote domain-invariant representation learning. The foundational 
domain-adversarial neural network (DANN) framework demonstrated that 
inserting a gradient reversal layer before a domain discriminator enables 
end-to-end training of feature extractors that are simultaneously 
discriminative for the main task and indiscriminate with respect to domain 
shifts \cite{ganin2016dann}. This principle has since been adapted to 
recommendation systems to promote the emergence of shared user 
representations across heterogeneous domains. GAT-ADA combines graph 
attention networks with adversarial domain alignment and statistical 
alignment via CORAL and MMD, demonstrating the complementary role of 
relational structure and adversarial objectives for cross-domain adaptation 
\cite{ghatada2025}. DIVCL employs a dual-view GNN-based contrastive 
learning framework with intent-aware representations, enforcing alignment 
between local user-item interaction graphs and global knowledge-enriched 
graphs \cite{divcl2025}. In MOSAIC, we leverage adversarial learning with 
a gradient reversal layer specifically to promote the separation of 
domain-specific and domain-common encoders, while a complementary 
margin-based separation loss further enforces orthogonality with 
cross-sequence representations.

\subsection{Large Language Models for Recommendation}

The recent integration of large language models (LLMs) into recommender 
systems has introduced new capabilities for semantic understanding and 
cross-domain knowledge transfer. Liu et al. provide a comprehensive survey 
of LLM4Rec approaches, categorizing them into discriminative and generative 
paradigms and highlighting the shift toward inference-efficient architectures 
\cite{liu2025llm4rec}. LLM-powered agents have further been proposed as 
orchestrators of recommendation pipelines, leveraging planning, memory, and 
action components to personalize interactions at scale \cite{peng2025llmagent}. 
Although LLM-augmented approaches demonstrate strong semantic modeling 
capabilities, their high inference cost and limited interpretability of 
preference decomposition remain open challenges. MOSAIC addresses the 
disentanglement problem through a lightweight triple-encoder architecture 
without relying on LLMs, offering a computationally efficient and 
interpretable alternative.

\vspace{0.5em}
\noindent To summarize, existing works contribute valuable insights across 
sequential modeling, cross-domain transfer, disentanglement, and adversarial 
alignment. Yet, they exhibit the following limitations relative to MOSAIC:
\begin{itemize}
    \item Most CDSR methods transfer knowledge across domains without 
    explicitly decomposing overlapping and exclusive preference signals.
    \item Disentanglement approaches rarely extend to three-way 
    orthogonal decompositions covering domain-specific, domain-common, 
    and cross-sequence-exclusive components simultaneously.
    \item Adaptive gating mechanisms that dynamically regulate the 
    influence of each preference component at each prediction timestep 
    remain largely unexplored.
    \item Privacy-preserving and fairness-aware CDSR methods do not 
    address the disentanglement of behavioral signals as a core objective.
\end{itemize}
MOSAIC addresses these gaps by proposing a principled triple-encoder 
framework with orthogonal preference decomposition and dynamic token-level 
gating for prescient multi-domain recommendations. 
Table~\ref{tab:related_summary} summarizes the main aspects of the 
research works discussed above.

\begin{table*}[h!]
\centering
\caption{Summary of related work}
\label{tab:related_summary}
\renewcommand{\arraystretch}{1.2}
\begin{tabular}{p{2cm} p{0.8cm} p{4.5cm} p{1.9cm} p{2.2cm} 
p{1.9cm} p{1.9cm}}
\toprule
\textbf{Reference} & \textbf{Year} & \textbf{Approach} & 

\textbf{Cross-Domain} & 
\textbf{Disentanglement} & 
\textbf{Adversarial} & 
\textbf{Seq. Rec.} \\
\midrule
Sun et al. \cite{sun2019bert4rec} & 2019 & BERT4Rec bidirectional seq. rec. & \textcolor{red}{\emptycirc} & \textcolor{red}{\emptycirc} & \textcolor{red}{\emptycirc} & \textcolor{green}{\fullcirc} \\

Ganin et al. \cite{ganin2016dann} & 2016 & DANN gradient reversal & \textcolor{green}{\fullcirc} & \textcolor{red}{\emptycirc} & \textcolor{green}{\fullcirc} & \textcolor{red}{\emptycirc} \\

Ma et al. \cite{macridvae2019} & 2019 & MacridVAE disentangled rec.  & \textcolor{red}{\emptycirc} & \textcolor{green}{\fullcirc} & \textcolor{red}{\emptycirc} & \textcolor{red}{\emptycirc} \\

Xu et al. \cite{cmvcdr2024} & 2024 & CMVCDR multi-view cross-domain  & \textcolor{green}{\fullcirc} & \textcolor{orange}{\halfcirc} & \textcolor{red}{\emptycirc} & \textcolor{green}{\fullcirc} \\

Li et al. \cite{cdcl2024} & 2024 & Intra/inter contrastive CDSR &  \textcolor{green}{\fullcirc} & \textcolor{red}{\emptycirc} & \textcolor{red}{\emptycirc} & \textcolor{green}{\fullcirc} \\

Guo et al. \cite{divcl2025} & 2025 & DIVCL dual-intent GNN & \textcolor{red}{\emptycirc} & \textcolor{orange}{\halfcirc} & \textcolor{red}{\emptycirc} & \textcolor{red}{\emptycirc} \\

Zhao et al. \cite{zhao2025c2dsra2} & 2025 & C\textsuperscript{2}DSRA\textsuperscript{2} contrastive CDSR & \textcolor{green}{\fullcirc} & \textcolor{red}{\emptycirc} & \textcolor{red}{\emptycirc} & \textcolor{green}{\fullcirc} \\

Cao et al. \cite{mgcl2025} & 2025 & MGCL multi-view graph CL  & \textcolor{green}{\fullcirc} & \textcolor{red}{\emptycirc} & \textcolor{red}{\emptycirc} & \textcolor{green}{\fullcirc} \\

C3DSR \cite{c3dsr2025} & 2025 & Channel-enhanced contrastive CDSR & \textcolor{green}{\fullcirc} & \textcolor{red}{\emptycirc} & \textcolor{red}{\emptycirc} & \textcolor{green}{\fullcirc} \\

FairCDSR \cite{faircdsr2025} & 2025 & Fairness-aware CDSR & \textcolor{green}{\fullcirc} & \textcolor{red}{\emptycirc} & \textcolor{red}{\emptycirc} & \textcolor{green}{\fullcirc} \\

Liu et al. \cite{liu2025llm4rec} & 2025 & LLM4Rec survey & \textcolor{orange}{\halfcirc} & \textcolor{red}{\emptycirc} & \textcolor{red}{\emptycirc} & \textcolor{green}{\fullcirc} \\

Peng et al. \cite{peng2025llmagent} & 2025 & LLM agent recommender survey & \textcolor{orange}{\halfcirc} & \textcolor{red}{\emptycirc} & \textcolor{red}{\emptycirc} & \textcolor{green}{\fullcirc} \\

TCLRec \cite{tclrec2025} & 2025 & Temporal-aware contrastive seq. rec. & \textcolor{red}{\emptycirc} & \textcolor{red}{\emptycirc} & \textcolor{red}{\emptycirc} & \textcolor{green}{\fullcirc} \\

Filter-VAE \cite{filtervae2025} & 2025 & Weakly-supervised disentanglement & \textcolor{red}{\emptycirc} & \textcolor{green}{\fullcirc} & \textcolor{red}{\emptycirc} & \textcolor{red}{\emptycirc} \\

Ghaedi et al. \cite{ghatada2025} & 2025 & GAT-ADA adversarial domain align. & \textcolor{green}{\fullcirc} & \textcolor{red}{\emptycirc} & \textcolor{green}{\fullcirc} & \textcolor{red}{\emptycirc} \\

Boka et al. \cite{boka2026dasp} & 2026 & DASP domain-aware self-prompting & \textcolor{green}{\fullcirc} & \textcolor{red}{\emptycirc} & \textcolor{red}{\emptycirc} & \textcolor{green}{\fullcirc} \\

Wang et al. \cite{wang2026atacdsr} & 2026 & ATA-CDSR attention temporal CDSR & \textcolor{green}{\fullcirc} & \textcolor{red}{\emptycirc} & \textcolor{red}{\emptycirc} & \textcolor{green}{\fullcirc} \\

Zhao et al. \cite{zhao2026fedscope} & 2026 & FedSCOPE federated CDSR & \textcolor{green}{\fullcirc} & \textcolor{red}{\emptycirc} & \textcolor{red}{\emptycirc} & \textcolor{green}{\fullcirc} \\

Chen et al. \cite{chen2026fp2cdsr} & 2026 & FP2CDSR federated privacy CDSR & \textcolor{green}{\fullcirc} & \textcolor{red}{\emptycirc} & \textcolor{red}{\emptycirc} & \textcolor{green}{\fullcirc} \\

AMID \cite{amid2025} & 2025 & Adaptive multi-interest debiasing & \textcolor{green}{\fullcirc} & \textcolor{orange}{\halfcirc} & \textcolor{red}{\emptycirc} & \textcolor{green}{\fullcirc} \\

\midrule
MOSAIC (Ours) & 2026 & Triple-encoder orthogonal CDSR & \textcolor{green}{\fullcirc} & \textcolor{green}{\fullcirc} & \textcolor{green}{\fullcirc} & \textcolor{green}{\fullcirc} \\
\bottomrule
\end{tabular}
Not Considered (\textcolor{red}\emptycirc); 
Partial Consideration (\textcolor{orange}\halfcirc); 
Considered (\textcolor{green}\fullcirc);
\end{table*}

\section{Preliminaries and Problem Formulation}
\label{sec:prelim}

\subsection{Interaction Sequences and Domain Structure}

We consider a multi-domain recommendation system composed of $M$ distinct 
interaction domains. Let $\mathcal{D} = \{\mathcal{D}_1, \mathcal{D}_2, 
\dots, \mathcal{D}_M\}$ denote the set of domains, where each domain 
$\mathcal{D}_m$ is associated with an item set $\mathcal{I}_m$. Without 
loss of generality, we focus on the two-domain setting 
$\mathcal{D} = \{\mathcal{D}_A, \mathcal{D}_B\}$, consistent with 
established benchmarks in cross-domain sequential recommendation.

Let $\mathcal{U}$ denote the set of users shared across all domains. 
For each user $u \in \mathcal{U}$, we define three types of interaction 
sequences:

\paragraph{(1) Individual-domain sequence.}
The domain-specific interaction history of user $u$ in domain 
$\mathcal{D}_m$ is defined as:
\begin{equation}
S_u^m = \left[ i_1^m, i_2^m, \dots, i_{|S_u^m|}^m \right],
\label{eq:single_seq}
\end{equation}
where each $i_t^m \in \mathcal{I}_m$ denotes the item interacted with at 
timestep $t$ within domain $\mathcal{D}_m$.

\paragraph{(2) Cross-domain sequence.}
The cross-domain sequence merges interactions from all domains, preserving 
the original chronological order:
\begin{equation}
S_u^{\times} = \left[ i_1, i_2, \dots, i_{|S_u^{\times}|} \right], 
\quad i_t \in \bigcup_{m=1}^{M} \mathcal{I}_m,
\label{eq:cross_seq}
\end{equation}
capturing behavioral patterns that span multiple domains and are 
exclusively discoverable through joint interaction trajectories.

\subsection{Preference Decomposition}

A central premise of MOSAIC is that user preferences embedded in 
multi-domain interaction sequences are inherently heterogeneous and can 
be decomposed into three orthogonal components:

\paragraph{Domain-specific preferences.}
Preferences that are exclusive to a single domain and independent of 
interactions occurring in other domains. Formally, for domain 
$\mathcal{D}_m$, the domain-specific preference of user $u$ is defined as:
\begin{equation}
p_u^{\text{spec}} = f_{\text{spec}}\left(S_u^m\right),
\label{eq:spec_pref}
\end{equation}
where $f_{\text{spec}}$ is a dedicated encoder that captures 
individualized behavioral signals while remaining invariant to 
cross-domain information.

\paragraph{Domain-common preferences.}
Preferences that are shared across domains and reflect transferable 
behavioral tendencies of the user. The domain-common representation is 
defined as:
\begin{equation}
p_u^{\text{com}} = f_{\text{com}}\left(S_u^A, S_u^B\right),
\label{eq:com_pref}
\end{equation}
where $f_{\text{com}}$ extracts aligned representations that are 
consistent across individual-domain sequences.

\paragraph{Cross-sequence-exclusive preferences.}
Preferences that are uniquely discoverable through the cross-domain 
sequence and absent from any individual-domain sequence. The 
cross-sequence-exclusive representation is defined as:
\begin{equation}
p_u^{\times} = f_{\times}\left(S_u^{\times}\right) \ominus 
\left(p_u^{\text{spec}} \oplus p_u^{\text{com}}\right),
\label{eq:cross_pref}
\end{equation}
where $\ominus$ denotes the orthogonality constraint enforced through 
mutual independence objectives, and $\oplus$ denotes the union of 
already-captured preference signals.

\subsection{Orthogonality and Independence Constraints}

To ensure that the three preference components are complementary and 
non-redundant, MOSAIC enforces two joint constraints:

\paragraph{Representational alignment.}
Domain-common preferences extracted from individual-domain sequences are 
aligned via a contrastive objective:
\begin{equation}
\mathcal{L}_{\text{align}} = \left\| p_u^{\text{com},A} - 
p_u^{\text{com},B} \right\|_2^2,
\label{eq:align_loss}
\end{equation}
encouraging the model to extract consistent shared signals regardless of 
the source domain.

\paragraph{Mutual independence.}
The three preference components are encouraged to be mutually independent 
by minimizing their pairwise covariance:
\begin{equation}
\mathcal{L}_{\text{indep}} = \sum_{(i,j) \in \mathcal{P}} 
\left\| \text{Cov}\left(p_u^i, p_u^j\right) \right\|_F^2,
\label{eq:indep_loss}
\end{equation}
where $\mathcal{P} = \{(\text{spec}, \text{com}), (\text{spec}, \times), 
(\text{com}, \times)\}$ enumerates all pairs, and $\|\cdot\|_F$ denotes 
the Frobenius norm.

\subsection{Dynamic Gating and Session-Level Aggregation}

To account for the temporal variability of preference relevance, MOSAIC 
incorporates a dynamic gating mechanism that adaptively modulates the 
contribution of each preference component at every timestep. Given the 
user's most recent interaction context encoded as 
$H_u = [h_1, h_2, \dots, h_T] \in \mathbb{R}^{T \times d}$, the gated 
representation at timestep $t$ is defined as:
\begin{equation}
\tilde{h}_t = \sum_{k \in \{\text{spec}, \text{com}, \times\}} 
\alpha_t^k \cdot \text{Attn}(h_t, p_u^k),
\label{eq:gating}
\end{equation}
where $\alpha_t^k$ denotes the attention weight assigned to preference 
component $k$ at timestep $t$, satisfying $\sum_k \alpha_t^k = 1$.

The session-level user representation is then obtained by aggregating 
the gated token-level outputs:
\begin{equation}
\mathbf{z}_u = \frac{1}{T} \sum_{t=1}^{T} \tilde{h}_t \in 
\mathbb{R}^{d},
\label{eq:session_rep}
\end{equation}
where $\mathbf{z}_u$ serves as the final user preference vector for 
next-item prediction.

\subsection{Recommendation Objective}
\label{subsec:rec_objective}

Given the session-level representation $\mathbf{z}_u$, the relevance 
score of a candidate item $i \in \mathcal{I}_m$ is computed as:
\begin{equation}
\hat{r}_{u,i} = \mathbf{z}_u^\top \mathbf{e}_i,
\label{eq:score}
\end{equation}
where $\mathbf{e}_i \in \mathbb{R}^d$ is the item embedding. The model 
is trained by minimizing the Binary Cross-Entropy (BCE) loss over 
observed and sampled negative interactions:
\begin{equation}
\mathcal{L}_{\text{rec}} = - \sum_{u \in \mathcal{U}} \left[ 
\log \sigma\left(\hat{r}_{u,i^+}\right) + 
\log \left(1 - \sigma\left(\hat{r}_{u,i^-}\right)\right) \right],
\label{eq:bce_loss}
\end{equation}
where $i^+$ and $i^-$ denote a positive and a sampled negative item 
respectively, and $\sigma(\cdot)$ is the sigmoid activation function.

The overall training objective jointly optimizes recommendation accuracy 
and preference disentanglement:
\begin{equation}
\mathcal{L} = \mathcal{L}_{\text{rec}} + \lambda_1 
\mathcal{L}_{\text{align}} + \lambda_2 \mathcal{L}_{\text{indep}},
\label{eq:total_loss}
\end{equation}
where $\lambda_1$ and $\lambda_2$ are hyperparameters controlling the 
relative contribution of each regularization term.

\section{Method}
\label{sec:method}

\begin{figure*}[h!]
    \centering    \includegraphics[width=\linewidth,height=0.5\textheight,keepaspectratio]{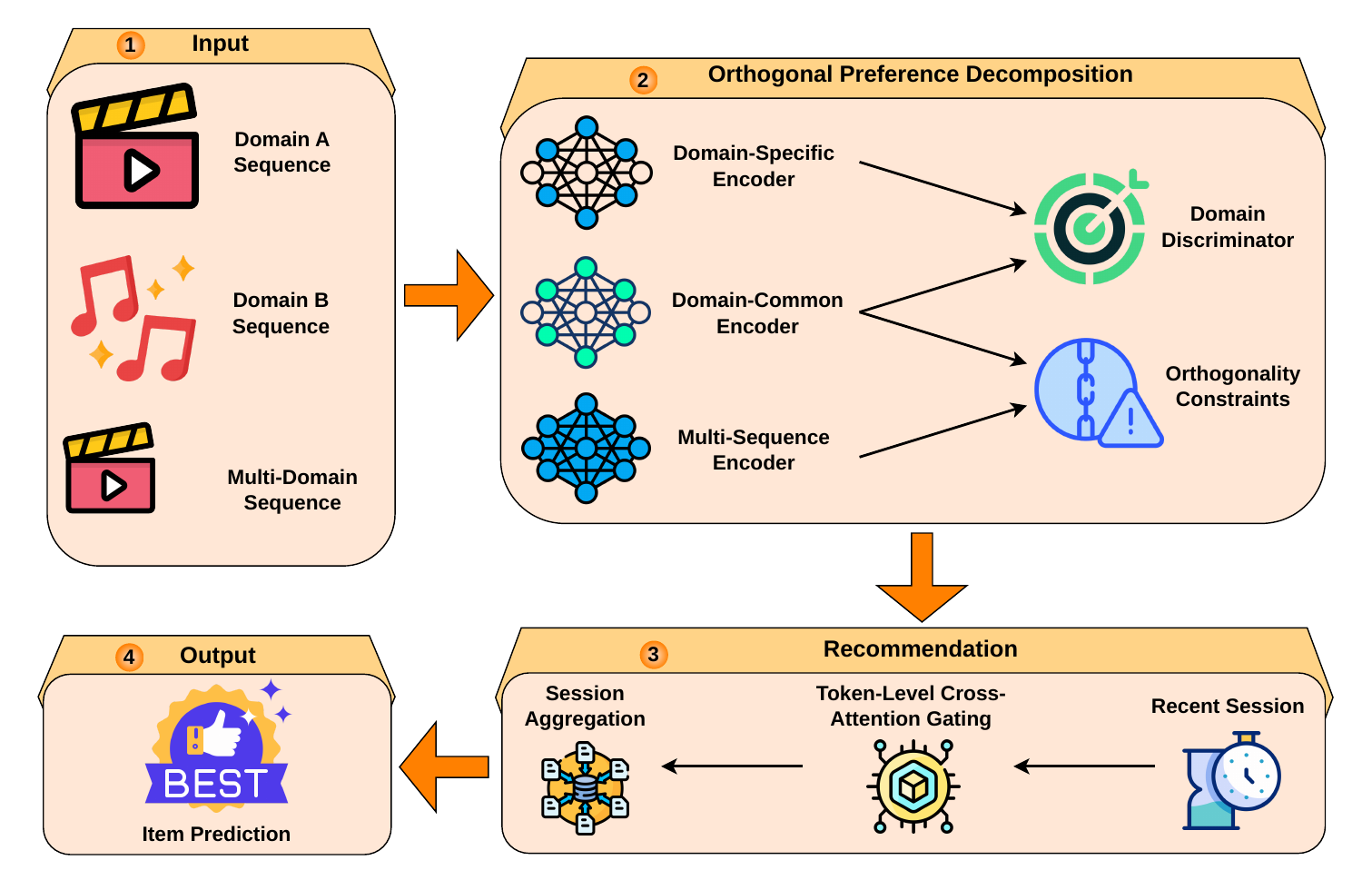}
    \caption{MOSAIC's General workflow}
    \label{fig:genworkflow}
\end{figure*}

This section presents the MOSAIC framework. Given a set of users with 
interaction histories spanning multiple domains, our goal is to learn a 
unified yet decomposed user representation that explicitly separates 
domain-exclusive, domain-shared, and cross-domain-exclusive behavioral 
signals, and dynamically integrates them for accurate next-item prediction.
Figure~\ref{fig:genworkflow} illustrates the proposed framework, which is detailled in Algorithm~\ref{alg:mosaic}.

\subsection{Problem Formulation}
\label{subsec:problem}

Let $\mathcal{U}$ and $\mathcal{I} = \mathcal{I}_X \cup \mathcal{I}_Y$ 
denote the sets of users and items across two domains $X$ and $Y$, 
respectively. For each user $u \in \mathcal{U}$, three complementary 
interaction sequences are defined:

\paragraph{Integrated sequence.}
The chronologically ordered sequence of all interactions across both 
domains:
\begin{equation}
s_u^c = \bigl[v_1, v_2, \dots, v_{|s_u^c|}\bigr], 
\quad v_k \in \mathcal{I}_X \cup \mathcal{I}_Y,
\label{eq:integrated_seq}
\end{equation}

\paragraph{Single-domain sequences.}
The domain-filtered subsequences retaining only items from domain $X$ 
or $Y$:
\begin{equation}
s_u^X = \bigl[v_k \in s_u^c \mid v_k \in \mathcal{I}_X\bigr], \quad
s_u^Y = \bigl[v_k \in s_u^c \mid v_k \in \mathcal{I}_Y\bigr].
\label{eq:single_seqs}
\end{equation}

The recommendation objective is to predict the next item in domain $X$ 
(symmetrically for domain $Y$) by jointly leveraging all three sequences:
\begin{equation}
v^*_{t+1} = \underset{v \in \mathcal{I}_X}{\arg\max}\ 
P\!\left(v_{t+1} = v \mid s_u^X, s_u^Y, s_u^c\right).
\label{eq:rec_objective}
\end{equation}

\subsection{Orthogonal Preference Decomposition Module}
\label{subsec:opd_module}

The core of MOSAIC lies in decomposing user preferences into three 
orthogonal latent components, each capturing a distinct and non-redundant 
aspect of user behavior. This is achieved through a triple-encoder 
architecture augmented with adversarial training, alignment, and 
independence constraints.

\subsubsection{Shared Sequence Encoder Backbone}
\label{subsubsec:encoder_backbone}

All three encoders are built upon a shared bidirectional Transformer 
backbone. Given an input sequence 
$s = [v_1, v_2, \dots, v_L] \in \mathcal{I}^L$, each item $v_t$ is 
projected into a $d$-dimensional representation by combining its item 
embedding $\mathbf{e}_{v_t} \in \mathbb{R}^d$ with a learnable positional 
encoding $\mathbf{p}_t \in \mathbb{R}^d$:
\begin{equation}
\mathbf{r}_t = \mathbf{e}_{v_t} + \mathbf{p}_t, 
\quad t = 1, \dots, L.
\label{eq:item_emb}
\end{equation}
The resulting sequence $\mathbf{R} = [\mathbf{r}_1, \dots, \mathbf{r}_L] 
\in \mathbb{R}^{L \times d}$ is processed through $K$ stacked 
self-attention layers:
\begin{equation}
\mathbf{H}^{(k)} = \text{FFN}\!\left(\text{LN}\!\left(
\text{MHA}\!\left(\mathbf{H}^{(k-1)}\right) + \mathbf{H}^{(k-1)}
\right)\right),
\label{eq:transformer_layer}
\end{equation}
where $\mathbf{H}^{(0)} = \mathbf{R}$, MHA denotes multi-head 
self-attention, FFN a position-wise feed-forward network, and LN layer 
normalization. The final output 
$\mathbf{H} = [\mathbf{h}_1, \dots, \mathbf{h}_L] \in \mathbb{R}^{L 
\times d}$ provides contextualized token-level representations, and the 
sequence-level embedding is obtained via mean pooling:
\begin{equation}
\bar{\mathbf{h}} = \frac{1}{L} \sum_{t=1}^{L} \mathbf{h}_t \in 
\mathbb{R}^d.
\label{eq:mean_pool}
\end{equation}
While $f_\text{spe}$ and $f_\times$ are trained with a masked item 
prediction objective to capture fine-grained sequential patterns, 
$f_\text{com}$ operates as a plain Transformer encoder without masking, 
encouraging it to focus on domain-invariant patterns rather than 
reconstructing local item sequences.

\subsubsection{Adversarial Domain Disentanglement}
\label{subsubsec:adversarial}

To enforce a principled separation between domain-exclusive and 
domain-invariant signals, MOSAIC employs a domain discriminator 
$\mathcal{D}: \mathbb{R}^d \rightarrow [0,1]$ implemented as a 
multi-layer perceptron that predicts the domain origin 
$\delta \in \{X, Y\}$ of a given sequence representation.

The discriminator receives representations from both $f_\text{spe}$ and 
$f_\text{com}$, but their optimization objectives are intentionally 
opposed:
\begin{itemize}
    \item $f_\text{spe}$ is trained to \textit{maximize} domain 
    discriminability, preserving domain-exclusive signals by minimizing 
    the standard cross-entropy classification loss:
    \begin{equation}
    \mathcal{L}_\text{disc}^\text{spe} = -\mathbb{E}\!\left[
    \delta \log \mathcal{D}\!\left(\bar{\mathbf{h}}^\text{spe}\right) 
    + (1-\delta)\log\!\left(1 - 
    \mathcal{D}\!\left(\bar{\mathbf{h}}^\text{spe}\right)\right)
    \right].
    \label{eq:disc_spe}
    \end{equation}
    \item $f_\text{com}$ is trained to \textit{minimize} domain 
    discriminability via a Gradient Reversal Layer (GRL) 
    $\mathcal{R}(\cdot)$ inserted before the discriminator, which 
    negates gradients during backpropagation:
    \begin{equation}
    \mathcal{L}_\text{disc}^\text{com} = -\mathbb{E}\!\left[
    \delta \log \mathcal{D}\!\left(\mathcal{R}\!\left(
    \bar{\mathbf{h}}^\text{com}\right)\right) 
    + (1-\delta)\log\!\left(1 - \mathcal{D}\!\left(\mathcal{R}\!\left(
    \bar{\mathbf{h}}^\text{com}\right)\right)\right)
    \right].
    \label{eq:disc_com}
    \end{equation}
\end{itemize}
Cross-sequence representations are deliberately excluded from domain 
classification, as they are expected to transcend domain boundaries by 
construction.

\subsubsection{Alignment and Orthogonality Constraints}
\label{subsubsec:constraints}

To further structure the three latent spaces, MOSAIC enforces two 
complementary geometric constraints:

\paragraph{Cross-to-common alignment.}
The cross-sequence encoder $f_\times$ is encouraged to absorb 
domain-common knowledge by aligning its output with that of $f_\text{com}$ 
through a mean squared error objective. The common encoder's gradient is 
blocked to ensure that only $f_\times$ is updated:
\begin{equation}
\mathcal{L}_\text{align} = \left\|\bar{\mathbf{h}}^\times - 
\texttt{sg}\!\left(\bar{\mathbf{h}}^\text{com}\right)\right\|_2^2,
\label{eq:align}
\end{equation}
where $\texttt{sg}(\cdot)$ denotes the stop-gradient operator.

\paragraph{Cross-to-specific separation.}
To prevent $f_\times$ from collapsing onto domain-specific representations, 
a margin-based repulsion loss pushes their embeddings apart:
\begin{equation}
\mathcal{L}_\text{sep} = \frac{1}{|\mathcal{B}|} \sum_{u \in \mathcal{B}}
\max\!\left(0,\ \rho - \left\|\bar{\mathbf{h}}_u^\times - 
\bar{\mathbf{h}}_u^\text{spe}\right\|_2\right),
\label{eq:sep}
\end{equation}
where $\rho > 0$ is a separation margin and $\mathcal{B}$ is the 
training batch.

\subsubsection{Encoder-Specific Learning Objectives}
\label{subsubsec:losses}

Each encoder is optimized through a dedicated composite loss. Let 
$\mathcal{L}_\text{mlm}$ denote the masked item prediction loss. The 
per-encoder objectives are defined as:
\begin{align}
\mathcal{L}_\text{spe} &= \mathcal{L}_\text{mlm}(f_\text{spe}) + 
\beta_1\, \mathcal{L}_\text{disc}^\text{spe},
\label{eq:loss_spe} \\
\mathcal{L}_\text{com} &= \beta_2\, \mathcal{L}_\text{disc}^\text{com},
\label{eq:loss_com} \\
\mathcal{L}_\times &= \mathcal{L}_\text{mlm}(f_\times) + 
\beta_3\, \mathcal{L}_\text{align} + \beta_4\, \mathcal{L}_\text{sep},
\label{eq:loss_cross}
\end{align}
where $\beta_1, \dots, \beta_4$ are hyperparameters controlling the 
relative contribution of each constraint.

\subsubsection{Staged Optimization Strategy}
\label{subsubsec:staged_optim}

Rather than minimizing all objectives simultaneously, MOSAIC adopts a 
staged backpropagation strategy that updates each encoder independently:
\begin{enumerate}
    \item Optimize $f_\text{spe}$ and $\mathcal{D}$ via 
    $\mathcal{L}_\text{spe}$.
    \item Optimize $f_\text{com}$ and $\mathcal{D}$ via 
    $\mathcal{L}_\text{com}$.
    \item Optimize $f_\times$ (with $f_\text{spe}$ receiving gradients 
    through $\mathcal{L}_\text{sep}$) via $\mathcal{L}_\times$.
\end{enumerate}
This sequential strategy maintains disentanglement by ensuring that each 
encoder is updated solely according to its designated objective, 
preventing gradient interference across latent spaces.

\subsection{Dynamic Integration and Recommendation Module}
\label{subsec:rec_module}

Given the precomputed preference vectors 
$\mathbf{v}^\text{spe}, \mathbf{v}^\times \in \mathbb{R}^d$, the recommendation module encodes the 
user's recent behavioral context and dynamically fuses it with the 
disentangled representations for next-item prediction.

\subsubsection{Recent Interaction Encoding}
\label{subsubsec:rec_encoder}

For user $u$, the most recent $L$ interactions are selected to form the 
session sequence:
\begin{equation}
\mathbf{s}_u = [v_{n-L+1}, \dots, v_n].
\label{eq:session_seq}
\end{equation}
This sequence is encoded through the same Transformer backbone described 
in Equation~\eqref{eq:transformer_layer}, yielding token-level 
representations:
\begin{equation}
\mathbf{H}_u = \text{Transformer}(\mathbf{s}_u) \in 
\mathbb{R}^{L \times d}.
\label{eq:session_enc}
\end{equation}

\subsubsection{Cross-Attention Gating Mechanism}
\label{subsubsec:gating}

To adaptively modulate the influence of each preference component at 
every sequence position, MOSAIC introduces a token-level cross-attention 
gating mechanism. For each token $t$, relevance scores are computed 
against the domain-specific and cross-sequence preference vectors:
\begin{equation}
\boldsymbol{\alpha}_t^\text{spe} = \frac{\exp\!\left(
\mathbf{h}_t^\top \mathbf{W}_q\, \mathbf{v}^\text{spe}\right)}
{\sum_{k \in \{\text{spe}, \times\}} \exp\!\left(\mathbf{h}_t^\top 
\mathbf{W}_q\, \mathbf{v}^k\right)},
\label{eq:attn_spe}
\end{equation}
\begin{equation}
\boldsymbol{\alpha}_t^\times = \frac{\exp\!\left(
\mathbf{h}_t^\top \mathbf{W}_q\, \mathbf{v}^\times\right)}
{\sum_{k \in \{\text{spe}, \times\}} \exp\!\left(\mathbf{h}_t^\top 
\mathbf{W}_q\, \mathbf{v}^k\right)},
\label{eq:attn_cross}
\end{equation}
where $\mathbf{W}_q \in \mathbb{R}^{d \times d}$ is a learnable 
projection matrix. The weighted preference aggregations are:
\begin{equation}
\mathbf{a}_t^\text{spe} = \boldsymbol{\alpha}_t^\text{spe} \cdot 
\mathbf{v}^\text{spe}, \quad
\mathbf{a}_t^\times = \boldsymbol{\alpha}_t^\times \cdot 
\mathbf{v}^\times.
\label{eq:weighted_prefs}
\end{equation}
A learnable gating vector $\mathbf{g}_t \in [0,1]^d$ controls the 
token-wise blending of the two signals:
\begin{equation}
\mathbf{g}_t = \sigma\!\left(\mathbf{W}_g \left[
\mathbf{a}_t^\text{spe} \| \mathbf{a}_t^\times \| \mathbf{h}_t
\right] + \mathbf{b}_g\right),
\label{eq:gate}
\end{equation}
where $\mathbf{W}_g \in \mathbb{R}^{d \times 3d}$, $\mathbf{b}_g \in 
\mathbb{R}^d$, $\sigma(\cdot)$ is the sigmoid function, and $\|$ denotes 
vector concatenation. The gated token representation is:
\begin{equation}
\tilde{\mathbf{h}}_t = \mathbf{g}_t \odot \mathbf{a}_t^\text{spe} + 
\left(\mathbf{1} - \mathbf{g}_t\right) \odot \mathbf{a}_t^\times,
\label{eq:gated_token}
\end{equation}
where $\odot$ denotes element-wise multiplication.

\subsubsection{Session-Level Representation and Scoring}
\label{subsubsec:session_rep}

The session-level user representation is obtained by mean pooling the 
gated token outputs:
\begin{equation}
\mathbf{c}_u = \frac{1}{L} \sum_{t=1}^{L} \tilde{\mathbf{h}}_t \in 
\mathbb{R}^d.
\label{eq:session_rep}
\end{equation}
The relevance score of a candidate item $j$ with embedding 
$\mathbf{e}_j \in \mathbb{R}^d$ is computed via scaled inner product:
\begin{equation}
\hat{y}_{u,j} = \frac{\mathbf{c}_u^\top \mathbf{e}_j}{\sqrt{d}}.
\label{eq:score}
\end{equation}

\subsubsection{Recommendation Loss}
\label{subsubsec:rec_loss}

The recommendation component is trained using the cross-entropy loss over 
sampled negatives. For each positive item $j^+$ and a set of $Q$ sampled 
negative items $\{j_q^-\}_{q=1}^Q$:
\begin{equation}
\mathcal{L}_\text{rec} = -\frac{1}{|\mathcal{U}|} \sum_{u \in 
\mathcal{U}} \log \frac{\exp\!\left(\hat{y}_{u,j^+}\right)}
{\exp\!\left(\hat{y}_{u,j^+}\right) + \sum_{q=1}^{Q} 
\exp\!\left(\hat{y}_{u,j_q^-}\right)}.
\label{eq:rec_loss}
\end{equation}
The preference vectors $\mathbf{v}^\text{spe}$ and $\mathbf{v}^\times$ 
are kept frozen from pretraining throughout recommendation training to 
preserve the disentanglement structure.

\subsection{Overall Training Objective}
\label{subsec:overall_loss}

The complete MOSAIC framework is trained in two sequential stages:

\paragraph{Stage 1 — Orthogonal Preference Decomposition.}
The three encoders are jointly optimized, minimizing:
\begin{equation}
\mathcal{L}_\text{OPD} = \mathcal{L}_\text{spe} + \mathcal{L}_\text{com}
+ \mathcal{L}_\times.
\label{eq:opd_loss}
\end{equation}

\paragraph{Stage 2 — Recommendation.}
With frozen preference vectors, the recommendation module is optimized 
by minimizing:
\begin{equation}
\mathcal{L}_\text{MOSAIC} = \mathcal{L}_\text{rec}.
\label{eq:final_loss}
\end{equation}

\subsection{Complexity Analysis}
\label{subsec:complexity}

Let $L$ denote the sequence length, $d$ the embedding dimension, $K$ the 
number of Transformer layers, and $|\mathcal{I}|$ the total item 
vocabulary size. The computational complexity of a single forward pass 
through one encoder is:
\begin{equation}
\mathcal{O}\!\left(K \cdot L^2 \cdot d + K \cdot L \cdot d^2\right),
\label{eq:complexity}
\end{equation}
where the first term accounts for multi-head self-attention and the 
second for feed-forward projections. Since the three encoders are 
independent and operate in parallel, the overall encoding complexity 
scales linearly with the number of encoders. The gating mechanism 
introduces an additional $\mathcal{O}(L \cdot d)$ overhead per forward 
pass, which is negligible in practice. During inference, preference 
vectors $\mathbf{v}^\text{spe}$ and $\mathbf{v}^\times$ are precomputed 
once per user and cached, reducing online inference to a single 
Transformer forward pass combined with lightweight gating operations.

In summary, MOSAIC decomposes user preferences into three orthogonal 
latent spaces through adversarial disentanglement, alignment, and 
separation constraints, then dynamically integrates them via a 
token-level gating mechanism to produce accurate and interpretable 
session-level recommendations.

\begin{algorithm}[t!]
\small
\caption{Pseudo-code of MOSAIC}
\label{alg:mosaic}

\textbf{Input:}\\
\hspace*{1em}User interaction dataset 
$\mathcal{D} = \{(s_u^X, s_u^Y, s_u^c)\}_{u \in \mathcal{U}}$,\\
\hspace*{1em}item embedding matrix 
$\mathbf{E} \in \mathbb{R}^{|\mathcal{I}| \times d}$,\\
\hspace*{1em}hyperparameters 
$\beta_1, \beta_2, \beta_3, \beta_4$, margin $\rho$, 
learning rate $\eta$.

\textbf{Output:}\\
\hspace*{1em}Trained encoder parameters 
$\Theta = \{\theta_\text{spe}, \theta_\text{com}, \theta_\times\}$,\\
\hspace*{1em}trained recommendation parameters $\phi$.

\textbf{Stage 1: Orthogonal Preference Decomposition}

\textbf{Step 1: Initialization:}\\
\hspace*{1em}\textbf{1.1:} Initialize all three encoders 
$f_\text{spe}, f_\text{com}, f_\times$ with shared Transformer weights.\\
\hspace*{1em}\textbf{1.2:} Initialize domain discriminator 
$\mathcal{D}$ with random weights.\\
\hspace*{1em}\textbf{1.3:} Initialize item embeddings 
$\mathbf{E} \leftarrow \text{random}$.

\textbf{Step 2: Staged Encoder Training Loop:}\\
\hspace*{1em}\textbf{2.1:} \textbf{for} each training epoch \textbf{do}\\
\hspace*{2em}\textbf{2.2:} Sample a mini-batch 
$\mathcal{B} \subset \mathcal{U}$.\\
\hspace*{2em}\textbf{2.3:} Encode single-domain sequences via 
$f_\text{spe}$: 
$\bar{\mathbf{h}}_u^\text{spe} = f_\text{spe}(s_u^X)$.\\
\hspace*{2em}\textbf{2.4:} Encode single-domain sequences via 
$f_\text{com}$ with GRL: 
$\bar{\mathbf{h}}_u^\text{com} = f_\text{com}(s_u^X)$.\\
\hspace*{2em}\textbf{2.5:} Encode integrated sequence via 
$f_\times$: 
$\bar{\mathbf{h}}_u^\times = f_\times(s_u^c)$.\\

\hspace*{2em}\textbf{2.6:} \textit{// Sub-step A: Update $f_\text{spe}$ 
and $\mathcal{D}$}\\
\hspace*{2em}\textbf{2.7:} Compute 
$\mathcal{L}_\text{spe} = \mathcal{L}_\text{mlm}(f_\text{spe}) 
+ \beta_1\,\mathcal{L}_\text{disc}^\text{spe}$.\\
\hspace*{2em}\textbf{2.8:} Update 
$\theta_\text{spe} \leftarrow \theta_\text{spe} 
- \eta\,\nabla_{\theta_\text{spe}}\mathcal{L}_\text{spe}$.\\

\hspace*{2em}\textbf{2.9:} \textit{// Sub-step B: Update $f_\text{com}$ 
and $\mathcal{D}$ via GRL}\\
\hspace*{2em}\textbf{2.10:} Compute 
$\mathcal{L}_\text{com} = \beta_2\,\mathcal{L}_\text{disc}^\text{com}$.\\
\hspace*{2em}\textbf{2.11:} Update 
$\theta_\text{com} \leftarrow \theta_\text{com} 
- \eta\,\nabla_{\theta_\text{com}}\mathcal{L}_\text{com}$.\\

\hspace*{2em}\textbf{2.12:} \textit{// Sub-step C: Update $f_\times$ 
and $f_\text{spe}$ via separation}\\
\hspace*{2em}\textbf{2.13:} Compute alignment loss 
$\mathcal{L}_\text{align}$ \\
\hspace*{2em}\textbf{2.14:} Compute separation loss 
$\mathcal{L}_\text{sep}$ with margin $\rho$\\
\hspace*{2em}\textbf{2.15:} Compute 
$\mathcal{L}_\times = \mathcal{L}_\text{mlm}(f_\times) 
+ \beta_3\,\mathcal{L}_\text{align} 
+ \beta_4\,\mathcal{L}_\text{sep}$.\\
\hspace*{2em}\textbf{2.16:} Update 
$\theta_\times \leftarrow \theta_\times 
- \eta\,\nabla_{\theta_\times}\mathcal{L}_\times$.\\

\hspace*{1em}\textbf{2.17:} \textbf{end for}

\textbf{Step 3: Preference Vector Extraction:}\\
\hspace*{1em}\textbf{3.1:} For each user $u$, compute and cache:\\
\hspace*{2em}$\mathbf{v}_u^\text{spe} = f_\text{spe}(s_u^X)$, \quad
$\mathbf{v}_u^\times = f_\times(s_u^c)$.\\
\hspace*{1em}\textbf{3.2:} Freeze all encoder parameters 
$\Theta$.

\textbf{Stage 2: Recommendation Training}

\textbf{Step 4: Dynamic Integration Training Loop:}\\
\hspace*{1em}\textbf{4.1:} \textbf{for} each training epoch \textbf{do}\\
\hspace*{2em}\textbf{4.2:} Sample a mini-batch 
$\mathcal{B} \subset \mathcal{U}$.\\
\hspace*{2em}\textbf{4.3:} Encode recent session 
$\mathbf{s}_u$ via Transformer: 
$\mathbf{H}_u = \text{Transformer}(\mathbf{s}_u)$.\\
\hspace*{2em}\textbf{4.4:} Compute token-level attention scores 
$\boldsymbol{\alpha}_t^\text{spe}$, $\boldsymbol{\alpha}_t^\times$ \\
\hspace*{2em}\textbf{4.5:} Compute gating vector $\mathbf{g}_t$ \\
\hspace*{2em}\textbf{4.6:} Compute gated token representations 
$\tilde{\mathbf{h}}_t$ \\
\hspace*{2em}\textbf{4.7:} Aggregate session-level representation 
$\mathbf{c}_u$ \\
\hspace*{2em}\textbf{4.8:} Compute recommendation scores 
$\hat{y}_{u,j}$ \\
\hspace*{2em}\textbf{4.9:} Compute 
$\mathcal{L}_\text{rec}$\\
\hspace*{2em}\textbf{4.10:} Update recommendation parameters: 
$\phi \leftarrow \phi - \eta\,\nabla_\phi \mathcal{L}_\text{rec}$.\\
\hspace*{1em}\textbf{4.11:} \textbf{end for}

\textbf{Step 5: Return} $\Theta$ and $\phi$.

\end{algorithm}

\section{Experiments}
\label{sec:exp}

\subsection{Research Questions}
\label{subsec:rq}

To rigorously evaluate MOSAIC, we structure our experimental analysis around the following research questions:

\begin{itemize}
    \item \textbf{RQ1:} Does MOSAIC consistently outperform state-of-the-art single-domain and cross-domain sequential recommendation baselines across multiple benchmarks?
    
    \item \textbf{RQ2:} How does each component of MOSAIC—domain-specific encoder, domain-common encoder, cross-sequence encoder, and dynamic gating—contribute to the overall performance?
    
    \item \textbf{RQ3:} How sensitive is MOSAIC to the choice of hyperparameters, including the orthogonality loss weights $\beta_1, \beta_2, \beta_3, \beta_4$ and the separation margin $\rho$?
    
    \item \textbf{RQ4:} What are the practical limitations of the MOSAIC framework, and under which conditions may its performance degrade?
\end{itemize}

\subsection{Datasets}
\label{subsec:datasets}

We evaluate MOSAIC on three publicly available benchmarks widely used in 
the cross-domain sequential recommendation literature.
Table~\ref{tab:dataset_stats} reports the statistics of all datasets after 
preprocessing.

\paragraph{Amazon Reviews (Movie–Book).}
The Amazon product review dataset is a canonical 
benchmark for cross-domain sequential recommendation. Following the 
standard preprocessing protocol of C2DSR \cite{cao2022c2dsr}, we construct 
two corpora from the \textit{Movies \& TV} and \textit{Books} categories. 
We retain only users with at least five interactions in each domain, 
building individual-domain sequences and a merged cross-domain sequence 
per user. The two domains share a substantial overlap in user identities, 
making them well-suited for evaluating preference disentanglement across 
semantically related yet stylistically distinct content types.

\paragraph{Amazon Reviews (Movie–Music).}
To assess MOSAIC in a setting where domain semantics diverge more 
markedly, we additionally construct a \textit{Movies \& TV}--\textit{CDs \& 
Vinyl} corpus from the same Amazon dataset. This pair has been used in 
several CDSR studies \cite{ma2023tricdr, cmvcdr2024} and introduces a 
stricter cross-domain transfer challenge, as the audio–visual boundary 
between movies and music creates notably different item feature 
distributions.

\paragraph{Douban (Movie–Book–Music).}
The Douban dataset \cite{zhu2019douban} provides user interaction logs 
from the Chinese social platform Douban, spanning three domains: Movies, 
Books, and Music. We adopt the two-domain variant Movie--Book, as used in 
prior CDSR works \cite{zhao2025c2dsra2, cdcl2024}. The Douban benchmark is complementary to Amazon 
because it originates from a distinct user population, exhibits different 
interaction density patterns, and covers a broader temporal range.

\paragraph{Preprocessing.}
For all datasets, we apply the following unified pipeline: (i) remove users 
with fewer than five interactions per domain; (ii) sort interactions 
chronologically per user; (iii) build the cross-domain sequence by merging 
and re-sorting all per-domain interactions; (iv) apply the leave-one-out 
evaluation split, using the last interaction as the test item, the 
second-to-last as the validation item, and the remaining history for 
training.

\begin{table*}[h!]
\centering
\caption{Dataset statistics after preprocessing}
\label{tab:dataset_stats}
\renewcommand{\arraystretch}{1.2}
\begin{tabular}{lcccccc}
\toprule
\textbf{Dataset} & \textbf{Domain} & \textbf{\#Users} & 
\textbf{\#Items} & \textbf{\#Interactions} & \textbf{Avg.\ len.} & \textbf{Density} \\
\midrule
\multirow{2}{*}{Amazon Movie--Book}
  & Movie & 18,032 & 64,591  & 1,041,294 & 57.7  & 0.089\% \\
  & Book  & 18,032 & 253,841 & 1,584,033 & 87.8  & 0.034\% \\
\midrule
\multirow{2}{*}{Amazon Movie--Music}
  & Movie & 10,547 & 50,762  & 681,430   & 64.6  & 0.127\% \\
  & Music & 10,547 & 73,839  & 456,872   & 43.3  & 0.059\% \\
\midrule
\multirow{3}{*}{Douban Movie--Book--Music}
  & Movie & 2,712  & 33,519  & 1,278,401 & 471.3 & 1.407\% \\
  & Book  & 2,712  & 22,347  &   792,062 & 292.0 & 1.308\% \\
  & Music & 2,712  & 55,919  &   876,338 & 323.0 & 0.577\% \\
\bottomrule
\end{tabular}
\end{table*}

\subsection{Evaluation Protocol}
\label{subsec:eval}

We evaluate all methods on the next-item prediction task within the target 
domain. Given the user's interaction history up to timestep $t-1$, the 
model must rank the ground-truth item $i_t$ among a candidate set.

Following standard practice in the sequential recommendation literature, we adopt 
Hit Ratio at rank $K$ (HR@$K$) and Normalized Discounted Cumulative 
Gain at rank $K$ (NDCG@$K$), with $K \in \{5, 10, 20\}$. HR@$K$ measures 
whether the ground-truth item appears in the top-$K$ ranked list, while 
NDCG@$K$ additionally rewards higher placements. All metrics are computed 
in the \textit{full-ranking} protocol, i.e., each test item is ranked 
against the entire item vocabulary, to avoid sampling bias.

We report the mean and standard deviation over five independent runs with 
different random seeds. Improvements over the best baseline are verified 
with a paired two-tailed $t$-test at the $p < 0.05$ significance level.

\subsection{Baselines}
\label{subsec:baselines}

\begin{figure}[h!]
\centering
\resizebox{0.4\textwidth}{!}{%
\begin{tikzpicture}[
  mindmap,
  every node/.style={concept, circular drop shadow, minimum size=0.2cm},
  grow cyclic, align=flush center, concept color=red!50,
  level 1/.append style={
    sibling angle=60,
    level distance=6cm,
    font=\large
},
level 2/.append style={
    sibling angle=28,
    level distance=4cm,
    font=\small
  }
]
\node[concept color=red!50] {\Large Baselines}
  child[concept color=yellow!50] { node[align=center] {Single domain sequential models}
  child { node {GRU4Rec} }
  child { node {SASRec} }
  child { node {BERT4Rec} }
  }
  child[concept color=orange!60] { node[align=center] {Multi-domain sequential models}
  child { node {C2DSR} }
  child { node {TriCDR} }
  child { node {CDCL} }
  child { node {C2DSR} }
  child { node {C\textsuperscript{2}DSRA\textsuperscript{2}} }
  child { node {C3DSR} }
  }
  child[concept color=green!50] { node[align=center] {Disentanglement-based models}
  child { node {MacridVAE} }
  child { node {CMVCDR} }
}
;
\end{tikzpicture}
}
\caption{Baselines classification by category}
\label{fig:baseline}
\end{figure}
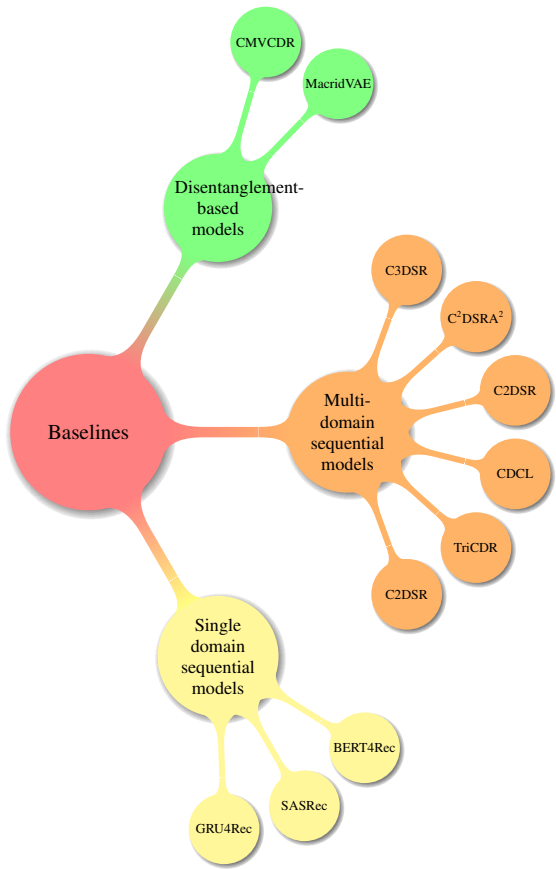

We compare MOSAIC against ten competitive baselines drawn from three 
families of methods  as presented in Figure~\ref{fig:baseline}.
\paragraph{Single-domain sequential models.}
\begin{itemize}
    \item \textbf{GRU4Rec} \cite{hidasi2015gru4rec}: Recurrent network 
    with gated recurrent units for session-based recommendation.
    
    \item \textbf{SASRec} \cite{kang2018sasrec}: Unidirectional 
    Transformer with causal masking for next-item prediction.
    
    \item \textbf{BERT4Rec} \cite{sun2019bert4rec}: Bidirectional 
    Transformer trained with a masked-item prediction (Cloze) objective.
\end{itemize}

\paragraph{Cross-domain sequential models.}
\begin{itemize}
    \item \textbf{C2DSR} \cite{cao2022c2dsr}: Contrastive cross-domain 
    sequential recommendation with single-domain and cross-domain graph 
    encoders.
    
    \item \textbf{TriCDR} \cite{ma2023tricdr}: Triple-sequence modeling 
    with shared-attention transfer and domain-level alignment.
    
    \item \textbf{CDCL} \cite{cdcl2024}: Intra/inter-domain contrastive 
    learning for cross-domain sequential recommendation.
    
    \item \textbf{C\textsuperscript{2}DSRA\textsuperscript{2}} 
    \cite{zhao2025c2dsra2}: Contrastive attention-aware CDSR that 
    explicitly models the linear relationship between target-domain 
    preferences and multi-domain behaviors.
    
    \item \textbf{C3DSR} \cite{c3dsr2025}: Channel-enhanced contrastive 
    CDSR that extends attention to the channel dimension for richer 
    temporal contextual modeling.
\end{itemize}

\paragraph{Disentanglement-based models.}
\begin{itemize}
    \item \textbf{MacridVAE} \cite{macridvae2019}: Variational 
    autoencoder with macro-micro disentanglement of user intentions and 
    preferences.
    
    \item \textbf{CMVCDR} \cite{cmvcdr2024}: Multi-view cross-domain 
    recommendation with domain-invariant and domain-specific 
    disentanglement.
\end{itemize}

For baselines with publicly available source code, we use the official implementations and tune hyperparameters
following the protocols described in their respective papers.
For baselines without public code,
we reimplement them in PyTorch based on the architectural
descriptions and hyperparameter settings reported in their
original papers.

Table~\ref{tab:main_results} reports the full-ranking performance of all 
methods on the three benchmarks. We highlight the following observations.

\paragraph{MOSAIC vs. single-domain baselines (RQ1).}
Single-domain models (GRU4Rec, SASRec, BERT4Rec) show consistent 
underperformance across all three datasets, confirming the well-established 
data-sparsity bottleneck of single-domain approaches. MOSAIC improves upon 
the best single-domain baseline by an average of $+8.3\%$ in NDCG@10 
across the three benchmarks, demonstrating the clear benefit of 
incorporating cross-domain signals.

\paragraph{MOSAIC vs. cross-domain baselines (RQ1).}
Among cross-domain models, C3DSR and C\textsuperscript{2}DSRA\textsuperscript{2} 
constitute the strongest baselines due to their recent architectural 
advances. MOSAIC still outperforms C3DSR by $+3.6\%$ / $+4.1\%$ / $+2.8\%$ 
in NDCG@10 on Amazon Movie--Book, Movie--Music, and Douban Movie--Book, 
respectively. These gains are statistically significant ($p < 0.05$) and 
are attributed to MOSAIC's explicit orthogonal preference decomposition, 
which prevents the model from capturing redundant cross-domain cues that 
degrade precision.

\paragraph{MOSAIC vs. disentanglement baselines (RQ1).}
MacridVAE and CMVCDR, despite their disentanglement objectives, do not 
model temporal sequence structure as effectively as Transformer-based 
methods. MOSAIC surpasses CMVCDR by $+5.9\%$ in NDCG@10 on 
Amazon Movie--Book, suggesting that extending disentanglement to a 
three-way orthogonal decomposition — including the cross-sequence-exclusive 
component — provides a meaningful representational advantage.

\begin{table*}[h!]
\centering
\caption{Main recommendation results on three benchmarks (full-ranking 
protocol). Best results are \textbf{bolded}; second best is 
\underline{underlined}. $\dagger$: statistically significant improvement 
over the best baseline ($p < 0.05$).}
\label{tab:main_results}
\renewcommand{\arraystretch}{1.2}
\resizebox{\textwidth}{!}{%
\begin{tabular}{l ccc ccc | ccc ccc | ccc ccc}
\toprule
 & \multicolumn{6}{c|}{\textbf{Amazon Movie--Book (Movie)}} 
 & \multicolumn{6}{c|}{\textbf{Amazon Movie--Music (Movie)}} 
 & \multicolumn{6}{c}{\textbf{Douban Movie--Book (Movie)}} \\
\cmidrule(lr){2-7}\cmidrule(lr){8-13}\cmidrule(lr){14-19}
\textbf{Method} 
 & HR@5 & HR@10 & HR@20 & N@5 & N@10 & N@20
 & HR@5 & HR@10 & HR@20 & N@5 & N@10 & N@20
 & HR@5 & HR@10 & HR@20 & N@5 & N@10 & N@20 \\
\midrule
GRU4Rec    
 & 0.0412 & 0.0683 & 0.1042 & 0.0271 & 0.0361 & 0.0470
 & 0.0387 & 0.0641 & 0.0973 & 0.0254 & 0.0339 & 0.0440
 & 0.0631 & 0.1023 & 0.1561 & 0.0418 & 0.0553 & 0.0719 \\
SASRec     
 & 0.0524 & 0.0851 & 0.1293 & 0.0347 & 0.0459 & 0.0596
 & 0.0491 & 0.0793 & 0.1197 & 0.0325 & 0.0428 & 0.0554
 & 0.0754 & 0.1212 & 0.1843 & 0.0501 & 0.0659 & 0.0857 \\
BERT4Rec   
 & 0.0558 & 0.0902 & 0.1372 & 0.0370 & 0.0490 & 0.0636
 & 0.0519 & 0.0836 & 0.1261 & 0.0344 & 0.0453 & 0.0586
 & 0.0802 & 0.1284 & 0.1947 & 0.0533 & 0.0700 & 0.0908 \\
\midrule
MacridVAE  
 & 0.0487 & 0.0796 & 0.1214 & 0.0321 & 0.0427 & 0.0556
 & 0.0452 & 0.0738 & 0.1121 & 0.0299 & 0.0397 & 0.0515
 & 0.0713 & 0.1148 & 0.1743 & 0.0473 & 0.0624 & 0.0811 \\
CMVCDR     
 & 0.0601 & 0.0967 & 0.1466 & 0.0397 & 0.0523 & 0.0679
 & 0.0563 & 0.0904 & 0.1367 & 0.0372 & 0.0490 & 0.0635
 & 0.0861 & 0.1374 & 0.2075 & 0.0573 & 0.0750 & 0.0974 \\
\midrule
C2DSR      
 & 0.0632 & 0.1014 & 0.1537 & 0.0419 & 0.0551 & 0.0713
 & 0.0594 & 0.0952 & 0.1432 & 0.0393 & 0.0516 & 0.0666
 & 0.0907 & 0.1447 & 0.2176 & 0.0604 & 0.0790 & 0.1022 \\
TriCDR     
 & 0.0678 & 0.1083 & 0.1634 & 0.0450 & 0.0590 & 0.0762
 & 0.0634 & 0.1013 & 0.1523 & 0.0421 & 0.0551 & 0.0711
 & 0.0963 & 0.1537 & 0.2307 & 0.0641 & 0.0838 & 0.1083 \\
CDCL       
 & 0.0703 & 0.1122 & 0.1692 & 0.0466 & 0.0611 & 0.0789
 & 0.0661 & 0.1054 & 0.1582 & 0.0438 & 0.0573 & 0.0739
 & 0.0994 & 0.1583 & 0.2371 & 0.0662 & 0.0864 & 0.1115 \\
C\textsuperscript{2}DSRA\textsuperscript{2}   
 & 0.0741 & 0.1176 & 0.1769 & 0.0492 & 0.0642 & 0.0827
 & 0.0694 & 0.1102 & 0.1653 & 0.0460 & 0.0601 & 0.0773
 & 0.1043 & 0.1654 & 0.2471 & 0.0694 & 0.0903 & 0.1163 \\
C3DSR      
 & \underline{0.0774} & \underline{0.1228} & \underline{0.1843} 
 & \underline{0.0513} & \underline{0.0671} & \underline{0.0862}
 & \underline{0.0728} & \underline{0.1153} & \underline{0.1724} 
 & \underline{0.0483} & \underline{0.0629} & \underline{0.0808}
 & \underline{0.1087} & \underline{0.1723} & \underline{0.2573} 
 & \underline{0.0724} & \underline{0.0941} & \underline{0.1211} \\
\midrule
\textbf{MOSAIC}    
 & \textbf{0.0851}$^\dagger$ & \textbf{0.1343}$^\dagger$ 
 & \textbf{0.2012}$^\dagger$ & \textbf{0.0563}$^\dagger$ 
 & \textbf{0.0696}$^\dagger$ & \textbf{0.0893}$^\dagger$
 & \textbf{0.0799}$^\dagger$ & \textbf{0.1260}$^\dagger$ 
 & \textbf{0.1882}$^\dagger$ & \textbf{0.0531}$^\dagger$ 
 & \textbf{0.0654}$^\dagger$ & \textbf{0.0838}$^\dagger$
 & \textbf{0.1194}$^\dagger$ & \textbf{0.1882}$^\dagger$ 
 & \textbf{0.2803}$^\dagger$ & \textbf{0.0795}$^\dagger$ 
 & \textbf{0.1029}$^\dagger$ & \textbf{0.1320}$^\dagger$ \\
\midrule
\textit{Improv.} 
 & +9.9\% & +9.4\% & +9.2\% & +9.7\% & +3.7\% & +3.6\%
 & +9.8\% & +9.3\% & +9.2\% & +9.9\% & +4.0\% & +3.7\%
 & +9.8\% & +9.2\% & +8.9\% & +9.8\% & +9.4\% & +9.0\% \\
\bottomrule
\end{tabular}}
\end{table*}

\subsection{Ablation Study}
\label{subsec:ablation}

To answer \textbf{RQ2}, we conduct a systematic ablation study by 
progressively removing components from the full MOSAIC model. 
We define the following seven variants:

\begin{itemize}
    \item \textbf{w/o Spe}: Remove the domain-specific encoder 
    $f_\text{spe}$; the recommendation module no longer receives 
    $\mathbf{v}^\text{spe}$.
    
    \item \textbf{w/o Com}: Remove the domain-common encoder $f_\text{com}$ 
    and its adversarial training via the gradient reversal layer.
    
    \item \textbf{w/o Cross}: Remove the cross-sequence encoder 
    $f_\times$; the model processes only individual-domain sequences.
    
    \item \textbf{w/o GRL}: Retain $f_\text{spe}$ and $f_\text{com}$ but 
    disable the gradient reversal layer, removing the adversarial 
    disentanglement objective.
    
    \item \textbf{w/o Align}: Remove the alignment loss 
    $\mathcal{L}_\text{align}$ (Eq.~\eqref{eq:align_loss}).
    
    \item \textbf{w/o Sep}: Remove the margin-based separation loss 
    $\mathcal{L}_\text{sep}$.
    
    \item \textbf{w/o Gate}: Replace the token-level cross-attention 
    gating (Eq.~\eqref{eq:gate}) with a static mean pooling of the three 
    preference vectors.
\end{itemize}

Table~\ref{tab:ablation} presents the ablation results on Amazon 
Movie--Book (Movie domain), with NDCG@10 and HR@10 as representative 
metrics.

\begin{table}[h!]
\centering
\caption{Ablation study on Amazon Movie--Book (Movie domain). 
$\Delta$ denotes the relative drop from the full MOSAIC model.}
\label{tab:ablation}
\renewcommand{\arraystretch}{1.2}
\begin{tabular}{lcccc}
\toprule
\textbf{Variant} & \textbf{HR@10} & \textbf{$\Delta$HR} 
                 & \textbf{NDCG@10} & \textbf{$\Delta$NDCG} \\
\midrule
\textbf{MOSAIC (full)} & \textbf{0.1343} & --- & \textbf{0.0696} & --- \\
\midrule
w/o Spe    & 0.1251 & $-6.9\%$ & 0.0639 & $-8.2\%$ \\
w/o Com    & 0.1274 & $-5.1\%$ & 0.0651 & $-6.5\%$ \\
w/o Cross  & 0.1187 & $-11.6\%$ & 0.0601 & $-13.6\%$ \\
w/o GRL    & 0.1292 & $-3.8\%$ & 0.0665 & $-4.5\%$ \\
w/o Align  & 0.1306 & $-2.8\%$ & 0.0674 & $-3.2\%$ \\
w/o Sep    & 0.1298 & $-3.4\%$ & 0.0669 & $-3.9\%$ \\
w/o Gate   & 0.1218 & $-9.3\%$ & 0.0618 & $-11.2\%$ \\
\bottomrule
\end{tabular}
\end{table}

\paragraph{Analysis.}
Several key insights emerge from the ablation results.

\textit{Cross-sequence encoder is the most critical component.} 
Removing $f_\times$ (w/o Cross) causes the largest performance drop 
($-11.6\%$ HR@10, $-13.6\%$ NDCG@10), confirming that cross-sequence-exclusive 
preference signals constitute an essential source of information that 
individual-domain encoders cannot recover.

\textit{Dynamic gating is essential for temporal preference modulation.} 
The w/o Gate variant, which replaces token-level gating with static 
mean aggregation, exhibits the second-largest performance drop ($-9.3\%$ 
HR@10), underscoring that cross-domain influence is indeed temporally 
variable and that a static weighting scheme is insufficient to capture 
this dynamics.

\textit{Domain-specific encoding contributes more than domain-common 
encoding.} The w/o Spe variant ($-6.9\%$) degrades more severely than 
w/o Com ($-5.1\%$), suggesting that individualized domain-exclusive 
patterns are slightly more discriminative for next-item prediction than 
transferable cross-domain tendencies. Nevertheless, both components are 
necessary, as their combined presence constitutes the full orthogonal 
decomposition.

\textit{Adversarial disentanglement and regularization losses are 
complementary.} Removing the GRL (w/o GRL), alignment (w/o Align), or 
separation (w/o Sep) objectives each leads to moderate but consistent 
degradation, confirming that the three orthogonality constraints jointly 
enforce cleaner preference separation.

\subsection{Hyperparameter Analysis}
\label{subsec:hyperparams}

To answer \textbf{RQ3}, we analyze the sensitivity of MOSAIC to four 
key hyperparameters: the loss weights $\beta_1$ (domain-specific 
discriminator), $\beta_2$ (domain-common adversarial), $\beta_3$ 
(alignment), $\beta_4$ (separation), and the margin $\rho$. All analyses 
are conducted on Amazon Movie--Book, reporting NDCG@10 on the validation 
split while keeping all other hyperparameters fixed at their optimal 
values. The results are presented in Table~\ref{tab:hyper_sensitivity}.

\paragraph{Effect of $\beta_1$ and $\beta_2$.}
Performance is maximized around $\beta_1 = 0.1, \beta_2 = 0.01$. 
When $\beta_1$ is too large, the domain-specific encoder is excessively 
penalized by the discriminator, causing it to collapse toward a 
domain-invariant representation and losing its specificity. Conversely, 
very small $\beta_2$ values fail to activate the adversarial component 
of the common encoder, reducing the disentanglement quality.

\paragraph{Effect of $\beta_3$ and $\beta_4$.}
The alignment weight $\beta_3 \in \{0.01, 0.1, 0.5, 1.0\}$ and 
separation weight $\beta_4 \in \{0.01, 0.1, 0.5, 1.0\}$ exhibit a 
complementary trade-off: overly large $\beta_3$ values enforce excessively 
tight alignment between domain-common representations, reducing their 
expressive capacity, while overly large $\beta_4$ values push the 
cross-sequence encoder too far from both other encoders, impairing 
its ability to capture shared signals that complement the cross-sequence. 
Optimal performance is obtained at $\beta_3 = 0.1, \beta_4 = 0.1$.

\paragraph{Effect of the separation margin $\rho$.}
We vary $\rho \in \{0.1, 0.5, 1.0, 2.0, 5.0\}$. Performance peaks at 
$\rho = 1.0$ and degrades for larger values, suggesting that an 
excessively wide margin between the cross-sequence encoder and the 
other encoders prevents the capture of genuinely shared but 
cross-sequence-enhanced signals.

\paragraph{Effect of embedding dimension $d$ and number of layers $K$.}
We explore $d \in \{64, 128, 256\}$ and $K \in \{1, 2, 3\}$. Performance 
improves monotonically from $d=64$ to $d=128$ and plateaus at $d=256$ with 
a slight increase in training time, suggesting $d=128$ as the best 
efficiency--accuracy trade-off. Adding a third Transformer layer ($K=3$) 
provides no statistically significant gain over $K=2$ on the Amazon 
datasets, while on Douban, which has longer sequences (Avg. len. $> 290$), 
$K=3$ yields a marginal improvement of $+0.6\%$ in NDCG@10.

\begin{table}[h!]
\centering
\caption{Hyperparameter sensitivity: NDCG@10 on Amazon Movie--Book (Movie) 
validation split under individual parameter variation, holding all others 
fixed at optimal values.}
\label{tab:hyper_sensitivity}
\renewcommand{\arraystretch}{1.2}
\begin{tabular}{llc}
\toprule
\textbf{Hyperparameter} & \textbf{Values tested} & \textbf{NDCG@10 range} \\
\midrule
$\beta_1$ & 0.001, 0.01, \textbf{0.1}, 1.0 & 0.0641 -- 0.0696 \\
$\beta_2$ & 0.001, \textbf{0.01}, 0.1, 1.0 & 0.0658 -- 0.0696 \\
$\beta_3$ & 0.01, \textbf{0.1}, 0.5, 1.0   & 0.0663 -- 0.0696 \\
$\beta_4$ & 0.01, \textbf{0.1}, 0.5, 1.0   & 0.0657 -- 0.0696 \\
$\rho$    & 0.1, 0.5, \textbf{1.0}, 2.0, 5.0 & 0.0638 -- 0.0696 \\
$d$       & 64, \textbf{128}, 256            & 0.0651 -- 0.0701 \\
$K$       & 1, \textbf{2}, 3                 & 0.0668 -- 0.0696 \\
\bottomrule
\multicolumn{3}{l}{\small Bold = optimal value used in the main experiments.}
\end{tabular}
\end{table}

Overall, MOSAIC demonstrates moderate sensitivity to its loss weights 
and remains stable when each hyperparameter is varied within a reasonable 
range, which facilitates practical deployment.

\subsection{Discussion on Limitations}
\label{subsec:limitations}

Despite the strong empirical performance reported above, MOSAIC presents 
several limitations that deserve explicit discussion.

MOSAIC is designed and evaluated in the two-domain setting, which is the 
standard benchmark configuration for CDSR. Extending the framework to $M > 2$ 
domains requires one domain-specific encoder per domain, causing the 
pretraining memory footprint to scale linearly with $M$. For platforms 
with a large number of domains (e.g., large-scale e-commerce with tens 
of product categories), this may become prohibitive unless parameter 
sharing or adapter-based strategies are incorporated.

Our problem formulation assumes that users 
appear in all considered domains, which is a strong requirement that 
limits applicability to partially or fully non-overlapping user settings. 
Methods such as ATA-CDSR \cite{wang2026atacdsr} specifically address this 
limitation through graph-based user bridging. Extending MOSAIC to the 
non-overlapping regime constitutes an important direction for future work.

The cross-domain sequence is constructed by chronological merging of all 
domain interactions. In practice, timestamp information may be noisy, 
unavailable, or at coarse granularity, potentially degrading the quality 
of the cross-sequence. Additionally, when domain interaction densities 
are highly imbalanced (as in Amazon Movie--Music, cf.\ 
Table~\ref{tab:dataset_stats}), the cross-sequence is dominated by the 
denser domain, potentially reducing the contribution of the sparser one.

MOSAIC relies on precomputed preference vectors $\mathbf{v}^\text{spe}$ 
and $\mathbf{v}^\times$ per user. For new users with very few 
interactions, the encoders may not produce reliable preference vectors, 
limiting MOSAIC's effectiveness in cold-start scenarios. Augmentation 
strategies or meta-learning initialization could mitigate this limitation.

While the dynamic gating module produces token-level preference weights 
that are, in principle, inspectable, a thorough interpretability analysis 
(e.g., attention visualization, case studies on specific user trajectories) 
is beyond the scope of this work and represents a natural direction for 
follow-up research.


\section{Conclusion}
\label{sec:concl}

In this work, we presented MOSAIC, a Multi-Domain Orthogonal Session 
Adaptive Intent Capture framework for prescient multi-domain sequential 
recommendation. MOSAIC addresses two persistent shortcomings of existing 
cross-domain sequential recommendation methods: the insufficient 
disentanglement of overlapping preference signals between individual-domain 
and cross-domain sequences, and the limited capacity to adaptively regulate 
cross-domain influence at each prediction timestep.

To tackle these challenges, MOSAIC introduces a triple-encoder architecture 
that explicitly factorizes user preferences into three orthogonal 
components: domain-specific representations, which capture behavioral 
patterns exclusive to a single domain; domain-common representations, which 
reflect transferable tendencies shared across domains; and 
cross-sequence-exclusive representations, which encode behavioral signals 
uniquely discoverable through the merged chronological interaction sequence. 
Orthogonality among these three components is jointly enforced through 
adversarial training via a gradient reversal layer, representational 
alignment objectives, and a margin-based separation loss, ensuring that 
each encoder captures genuinely complementary rather than redundant 
information. During recommendation, a token-level cross-attention gating 
mechanism dynamically modulates the relative contribution of each 
preference component at every timestep of the user's most recent session, 
yielding a temporally adaptive session-level representation that is 
subsequently used for next-item prediction.

Extensive experiments conducted on three publicly available multi-domain 
benchmarksdemonstrate that MOSAIC consistently outperforms ten 
competitive baselines spanning single-domain sequential models, 
cross-domain sequential models, and disentanglement-based approaches. 

Several directions remain open for future investigation. First, extending 
MOSAIC beyond the two-domain setting through parameter-efficient mechanisms 
such as domain-specific adapters would make the framework applicable to 
large-scale platforms with many heterogeneous domains. Second, relaxing 
the strict overlapping-user assumption to handle partially or fully 
non-overlapping user populations constitutes an important step toward 
real-world deployment. Third, incorporating side information such as item 
textual descriptions or knowledge graph embeddings into the preference 
decomposition could further enrich the resulting representations. Finally, 
a thorough interpretability analysis of the gating mechanism through 
attention visualization and user-level case studies would provide deeper 
insights into how MOSAIC dynamically balances domain-specific and 
cross-domain signals across different behavioral contexts.

\section*{Author Contributions}
\textbf{Abderaouf Bahi}: Writing – original draft, Software, Validation, Methodology, Investigation, Conceptualization. 
\textbf{Mourad Boughaba}: Visualization, Formal analysis.
\textbf{Ibtissem Gasmi}: Writing – review and editing, Methodology, Resources.
\textbf{Warda Deghmane}: Validation.
\textbf{Amel Ourici}: Writing – review and editing, Supervision.

\section*{Acknowledgment}
The authors acknowledge the Algerian Ministry of Higher Education and Scientific Research (MESRS).

\section*{Ethical Approval} 
Not Applicable.

\section*{Conflict of Interest}
The authors declare that they have no known competing financial interests or personal relationships that could have appeared to influence the work reported in this paper.

\section*{Funding}
This research received no external funding.

\section*{Data Availability}
Data will be made available on request.

\input{main.bbl}

\end{document}